# Tensor damping in metallic magnetic multilayers

Neil Smith

San Jose Research Center, Hitachi Global Storage Technologies, San Jose, CA 95135

The mechanism of spin-pumping, described by Tserkovnyak *et al*. , is formally analyzed in the general case of a magnetic multilayer consisting of two or more metallic ferromagnetic (FM) films separated by normal metal (NM) layers. It is shown that the spin-pumping-induced dynamic coupling between FM layers modifies the linearized Gilbert equations in a way that replaces the scalar Gilbert damping constant with a nonlocal matrix of Cartesian damping tensors. The latter are shown to be methodically calculable from a matrix algebra solution of the Valet-Fert transport equations. As an example, explicit analytical results are obtained for a 5-layer (spin-valve) of form NM/FM/NM'/FM/NM. Comparisons with earlier well known results of Tserkovnyak *et al*. for the related 3-layer FM/NM/FM indicate that the latter inadvertently hid the tensor character of the damping, and instead singled out the diagonal element of the local damping tensor along the axis normal to the plane of the two magnetization vectors. For spin-valve devices of technological interest, the influence of the tensor components of the damping on thermal noise or spin-torque critical currents are strongly weighted by the relative magnitude of the elements of the nonlocal, anisotropic stiffness-field tensor-matrix, and for in-plane magnetized spin-valves are generally more sensitive to the in-plane element of the damping tensor.

## I. INTRODUCTION

For purely scientific reasons, as well as technological applications such as magnetic field sensors or dc current tunable microwave oscillators, there is significant present interest[1] in the magnetization dynamics in current-perpendicular-to-plane (CPP) metallic multilayer devices comprising multiple ferromagnetic (FM) films separated by normal metal (NM) spacer layers. The phenomenon of spin-pumping, described earlier by Tserkovnyak et al.[2,3] introduces an additional source of dynamic coupling, either between the magnetization of a single FM layer and its NM electronic environment, or between two or more FM layers as mediated through their NM spacers. In the former case,[2] the effect can resemble an enhanced magnetic damping of an individual FM layer, which has important practical application for substantially increasing the spin-torque critical currents of CPP spin-valves employed as giant-magnetoresistive (GMR) sensors for read head applications.[4] Considered in this paper is a general treatment in the case of two or more FM layers in a CPP stack, where it will be shown in Sec. II that spin-pumping modifies the linearized equations of motion in a way that replaces a scalar damping constant with a nonlocal matrix of Cartesian damping tensors.[5] Analytical results for the case of a 5-layer spin-valve stack of the form NM/FM/NM'/FM/NM are discussed in detail in Sec. III, and are in Sec. IV compared and contrasted with the early well-known results of Tserkovnyak *et al.*.[3], as well as some very recent results of that author and colleagues.[6] In the case of CPP-GMR devices of technological interest, the relative importance of the different elements of the damping tensor on influencing measureable thermal fluctuations or spin-torque critical currents is shown to be strongly weighted by the anisotropic nature of the stiffness-field tensor-matrix.

## II. SPIN-PUMPING AND TENSOR DAMPING

As discussed by Tserkovnyak et al,[2,3] the spin current $\boldsymbol{I}^{\text{pump}}$ flowing into the normal metal (NM) layer at an FM/NM interface (Fig. 1) due to the spin-pumping effect is described the expression

$$\boldsymbol{I}^{\text{pump}} = \frac{\hbar}{4\pi}\left[\text{Re}\, g^{\uparrow\downarrow}(\hat{\boldsymbol{m}} \times \frac{d\hat{\boldsymbol{m}}}{dt}) - \text{Im}\, g^{\uparrow\downarrow}\frac{d\hat{\boldsymbol{m}}}{dt}\right] \tag{1}$$

where $g^{\uparrow\downarrow}$ is a dimensionless mixing conductance, and $\hat{\boldsymbol{m}}$ is the unit magnetization vector. In this paper, $\hat{\boldsymbol{m}}$ for any ferromagnetic (FM) layer is treated as a uniform macrospin. A restatement of (1) in terms more natural to Valet-Fert[7]form of transport equations is discussed in Appendix A. With the notational conversion $\boldsymbol{I}^{\text{pump}} \rightarrow -(\hbar/2e)A\boldsymbol{J}^{\text{pump}}$, where $A$ is the cross sectional area of the film stack, equation (1), for the case $\text{Re}\, g^{\uparrow\downarrow} >> \text{Im}\, g^{\uparrow\downarrow}$, simplifies to

$$\boldsymbol{J}^{\text{pump}} \cong \mp \frac{e}{2\pi}\frac{(h/2e^2)}{r^{\uparrow\downarrow}}\left(\hat{\boldsymbol{m}}\times\frac{d\hat{\boldsymbol{m}}}{dt}+\varepsilon\frac{d\hat{\boldsymbol{m}}}{dt}\right),\ \varepsilon\equiv\frac{\text{Im}\,r^{\uparrow\downarrow}}{\text{Re}\,r^{\uparrow\downarrow}} \qquad (2)$$

"−" for FM/NM interface, "+" for NM/FM interface

where $r^{\uparrow\downarrow} \cong \text{Re}\,r^{\uparrow\downarrow}$ is the inverse mixing conductance (with dimensions of resistance-area), and $h/2e^2$ is the well known inverse conductance quantum $(\cong 12.9\,\text{k}\Omega)$. In the present notation, all spin current densities $\boldsymbol{J}^{\text{spin}}$ have the same dimensions as electron charge current density $J_e$, and for conceptual simplicity are defined with a parallel (i.e., $\hat{\boldsymbol{J}}^{\text{spin}} = +\hat{\boldsymbol{m}}$) rather than anti-parallel alignment with magnetization $\hat{\boldsymbol{m}}$. Positive $J$ is defined as electrons flowing to the right (along $+\hat{\boldsymbol{y}}$ in Fig. 1.).

For a FM layer sandwiched by two NM layers in which the FM layer is the $j^{\text{th}}$ layer $(j \geq 0)$ of a

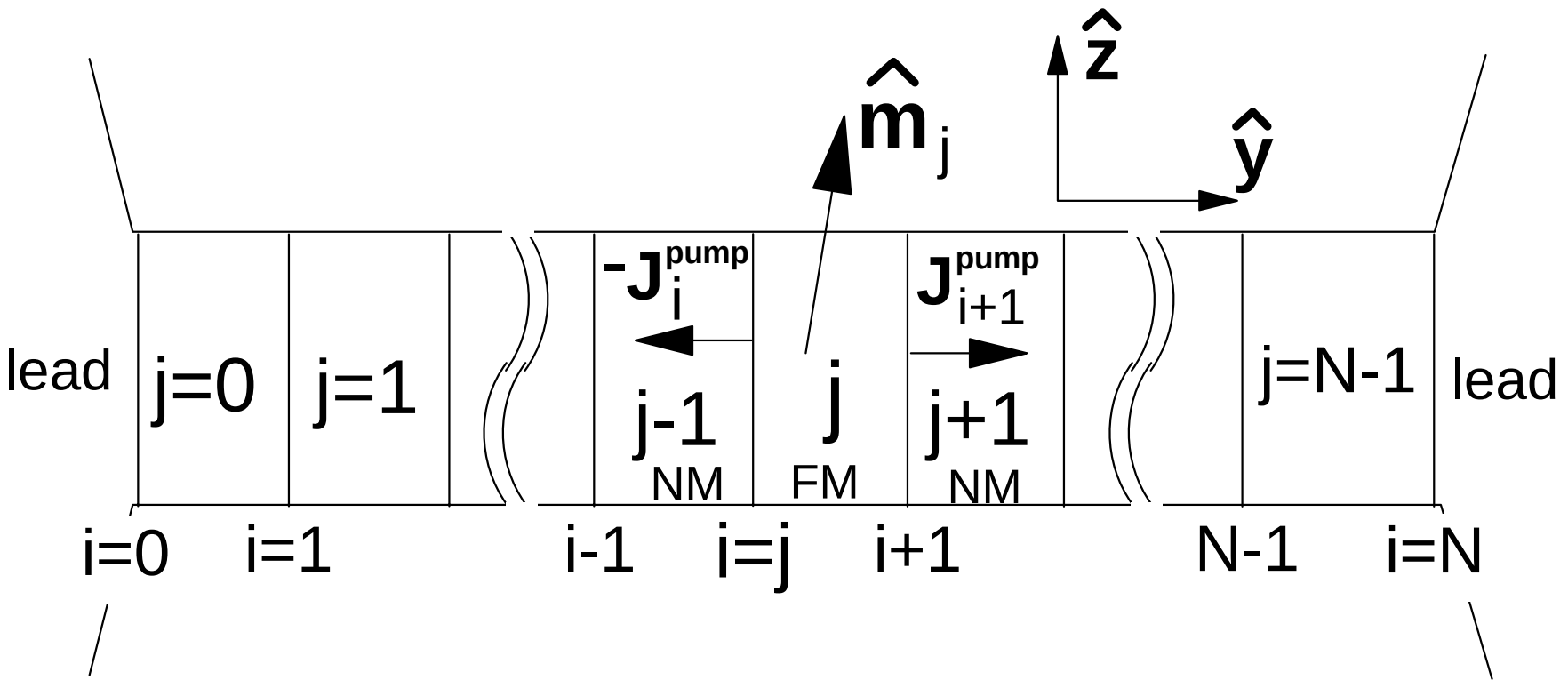

Fig. 1 Cross section cartoon of an *N*-layer multilayer stack with *N*-1 interior nterfaces of FM-NM or NM-NM type, such as found in CPP-GMR pillars sandwiched between conductive leads of much larger cross section. In the example shown, the $j^{\text{th}}$ layer is FM, sandwiched by NM layers, with spin pumping contributions at the $i^{\text{th}}$ (NM/FM) and $(i+1)^{\text{st}}$ (FM/NM) interfaces located at $y = y_i$ and $y_{i+1}$ (with $i=j$ for the labeling scheme shown).

multilayer

film stack, spin-pumping contributions at the $i^{\text{th}}$ interface, i.e., either left $(i = j)$ or right $(i = j+1)$ FM-NM interfaces, (2) can be expressed as

$$\boldsymbol{J}^{\text{pump}}_{i=j,j+1} = \frac{\hbar}{2e}\frac{(-1)^{i-j}}{r_i^{\uparrow\downarrow}}\left(\hat{\boldsymbol{m}}_j \times \frac{d\hat{\boldsymbol{m}}_j}{dt} + \varepsilon_i\frac{d\hat{\boldsymbol{m}}_j}{dt}\right) \qquad (3)$$

The physical picture to now be invoked is that of *small* (thermal) fluctuations of $\hat{\boldsymbol{m}}$ about equilibrium $\hat{\boldsymbol{m}}_0$ giving rise to the $d\hat{\boldsymbol{m}}/dt$ terms in (2). Since $|\hat{\boldsymbol{m}}| \equiv 1$, the three vector components of $\hat{\boldsymbol{m}}$ and/or $d\hat{\boldsymbol{m}}/dt$ are not linearly independent. To remove this interdependency, as well as higher order terms in (3) it thus is useful to work in a primed coordinate system where $\hat{\boldsymbol{z}}' = \hat{\boldsymbol{m}}_0$, through use of a $3\times 3$ Cartesian rotation matrix $\Re(\hat{\boldsymbol{m}}_0)$ such that $\hat{\boldsymbol{m}} = \Re \cdot \hat{\boldsymbol{m}}'$.[8] To *first* order in linearly *independent* quantities $(m'_x, m'_y)$, $\hat{\boldsymbol{m}} = \hat{\boldsymbol{m}}_0 + \tilde{\Re} \cdot \boldsymbol{m}'$, where $\boldsymbol{m}' \equiv \begin{pmatrix} m'_{x'} \\ m'_{y'} \end{pmatrix}$, and where $\tilde{\Re}$ denotes the $3\times 2$ matrix from the first two (i.e., *x* and *y*) columns of $\Re$. Replacing $\hat{\boldsymbol{m}}_0 \to \hat{\boldsymbol{z}}'$, and $\hat{\boldsymbol{z}}' \times _$ with matrix multiplication, the *linearized* form of (3) becomes

$$\boldsymbol{J}^{\text{pump}}_{i=j,j+1} = \frac{\hbar}{2e} \frac{(-1)^{i-j}}{r_i^{\uparrow\downarrow}} \tilde{\Re}_j \cdot \begin{pmatrix} \varepsilon_i & -1 \\ 1 & \varepsilon_i \end{pmatrix} \cdot \begin{pmatrix} dm'_{jx'}/dt \\ dm'_{jy'}/dt \end{pmatrix} \tag{4}$$

Using the present sign convention, $\boldsymbol{S}_j = (M_s t)_i A/\gamma\, \hat{\boldsymbol{m}}_j$ is the spin angular momentum of the $j^{\text{th}}$ FM layer with saturation magnetization-thickness product $(M_s t)_j$, and $\gamma > 0$ is the gyromagnetic ratio. Taking $|\boldsymbol{M}| = M_s$ as constant, it follows by angular momentum conservation that[3]

$$\frac{(M_s t)_j}{\gamma} \frac{d\hat{\boldsymbol{m}}_j}{dt} \Leftrightarrow \frac{1}{A} \frac{d\boldsymbol{S}_j}{dt} = \frac{\hbar}{2e} \sum_{i=j}^{j+1} (-1)^{i-j}\, \hat{\boldsymbol{m}}_j \times \boldsymbol{J}_i^{\text{NM}} \times \hat{\boldsymbol{m}}_j \tag{5}$$

is the contribution to $d\hat{\boldsymbol{m}}_j/dt$ due to the *net transverse* spin current entering the $j^{\text{th}}$ FM layer (Fig. 1). In (5), $\boldsymbol{J}_i^{\text{NM}}$ denotes the spin-current density *in* the NM layer *at* the $i^{\text{th}}$ FM-NM interface. Taking the cross product $\hat{\boldsymbol{m}}\times$ on both sides of (5), transforming to primed coordinates by matrix-multiplying by $\Re^{-1} = \Re^{\mathsf{T}}$, and employing similar linearization as to obtain (4), one finds to first order that

$$\hat{\boldsymbol{z}}'_j \times \frac{1}{A} \frac{d\boldsymbol{S}'_j}{dt} = \frac{\hbar}{2e} \begin{pmatrix} 0 & -1 \\ 1 & 0 \end{pmatrix} \cdot \tilde{\Re}_j^{\mathsf{T}} \cdot \left[ \Delta \boldsymbol{J}_j^{\text{spin}} \equiv \sum_{i=j}^{j+1} (-1)^{i-j} \boldsymbol{J}_i^{\text{NM}} \right] \tag{6}$$

where $\tilde{\Re}^{\mathsf{T}}$ is the $2\times 3$ matrix transpose of $\tilde{\Re}$. By definition, $\tilde{\Re}_j^{\mathsf{T}} \cdot \hat{\boldsymbol{m}}_{0j} = 0$.

The quantities $\Delta \boldsymbol{J}_j^{\text{spin}}$ in (6) are not known *a priori*, but must be determined after solution of the appropriate transport equations (e.g., Appendix B). Even in the absence of charge current flow (i.e., $J_e = 0$) as considered here, the $\Delta \boldsymbol{J}_j^{\text{spin}}$ are nonzero due to the set of $\boldsymbol{J}_i^{\text{pump}}$ in (4) which appear as source terms in the boundary conditions (A9) at each FM-NM interface. Given the *linear* relation of (4), one can now apply linear superposition to express

$$\Delta \boldsymbol{J}_j^{\mathrm{spin}} = \frac{\hbar}{2e}\sum_k \frac{1}{\bar{r}_k^{\uparrow\downarrow}} \overset{\leftrightarrow}{C}_{jk} \cdot \tilde{\Re}_k \cdot \begin{pmatrix} \varepsilon & -1 \\ 1 & \varepsilon \end{pmatrix} \frac{d\boldsymbol{m}'_k}{dt}, \quad \frac{1}{\bar{r}_k^{\uparrow\downarrow}} \equiv \frac{1}{2}\sum_{i=k}^{k+1} \frac{1}{r_i^{\uparrow\downarrow}} \tag{7}$$

in terms of the set of 3-D dimensionless Cartesian tensor $\overset{\leftrightarrow}{C}_{jk}$. The $\overset{\leftrightarrow}{C}_{jk}$ are convenient for formal expressions such as (9), or for analytical work in algebraically simple cases, such as exampled in Sec.III. However, they are also subject to methodical computation. For the kth magnetic layer, the $1^{\mathrm{st}}$, $2^{\mathrm{nd}}$, or $3^{\mathrm{rd}}$ columns of each $\overset{\leftrightarrow}{C}_{jk}$ are the dimensionless vectors $\Delta \boldsymbol{J}_j^{\mathrm{spin}}$ simultaneously obtainable for all magnetic layers $j$ from a matrix solution[9] of the Valet-Fert[7] transport equations with nonzero dimensionless spin-pump vectors $\boldsymbol{J}_{i=k,k+1}^{\mathrm{pump}} = (-1)^{i-k}(\bar{r}_k^{\uparrow\downarrow} / r_i^{\uparrow\downarrow})(\hat{\boldsymbol{x}}, \hat{\boldsymbol{y}}, \text{or } \hat{\boldsymbol{z}})$.

To include spin currents via (5) into the magnetization dynamics, the conventional Gilbert equations of motion for $\hat{\boldsymbol{m}}(t)$ can be amended as

$$\frac{d\hat{\boldsymbol{m}}_j}{dt} = -\gamma(\hat{\boldsymbol{m}}_i \times \boldsymbol{H}_j^{\mathrm{eff}}) + \alpha_j^{\mathrm{G}} \hat{\boldsymbol{m}}_j \times \frac{d\hat{\boldsymbol{m}}_j}{dt} + \frac{\gamma}{(M_s t)_j}\frac{1}{A}\frac{d\boldsymbol{S}_j}{dt} \tag{8}$$

where $\alpha_j^{\mathrm{G}}$ is the usual (scalar) Gilbert damping parameter. From (6) and (7), one can deduce that the rightmost term in (8) will scale linearly with $d\boldsymbol{m}'/dt$, as does the conventional Gilbert damping term. Combining these terms together after applying the analogous linearization procedure to (8) as was done in going from (5) to (6), one obtains

$$\hat{\boldsymbol{z}}'_j \times \frac{d\boldsymbol{m}'_j}{dt} = \gamma[\tilde{\Re}_j^{\mathrm{T}} \cdot \boldsymbol{H}_j^{\mathrm{eff}} - (\hat{\boldsymbol{m}}_j \cdot \boldsymbol{H}_j^{\mathrm{eff}})\boldsymbol{m}'_j] - \sum_k \overset{\leftrightarrow}{\alpha}'_{jk} \cdot \frac{d\boldsymbol{m}'_k}{dt}$$

$$\overset{\leftrightarrow}{\alpha}'_{jk} \equiv \begin{pmatrix} \alpha_j^{\mathrm{G}} & 0 \\ 0 & \alpha_j^{\mathrm{G}} \end{pmatrix} \delta_{jk} + \overset{\leftrightarrow}{\alpha}'^{\mathrm{pump}}_{jk} \tag{9}$$

$$\overset{\leftrightarrow}{\alpha}'^{\mathrm{pump}}_{jk} = \frac{\hbar\gamma}{(4\pi M_s t)_j}\frac{h/2e^2}{\bar{r}_j^{\uparrow\downarrow}} \begin{pmatrix} 0 & 1 \\ -1 & 0 \end{pmatrix} \cdot \tilde{\Re}_j^{\mathrm{T}} \cdot \overset{\leftrightarrow}{C}_{jk} \cdot \tilde{\Re}_k \cdot \begin{pmatrix} \varepsilon & -1 \\ 1 & \varepsilon \end{pmatrix}$$

where Kronecker delta $\delta_{jk} = 1$ if $j = k$, and $\delta_{jk} = 0$ if $j \neq k$.

In (9), $\overset{\leftrightarrow}{\alpha}'_{jj}$ is a 2-dimensional Cartesian "damping tensor" expressed in a coordinate system where $\hat{\boldsymbol{m}}'_{0j} = \hat{\boldsymbol{z}}'_j$, while $\overset{\leftrightarrow}{\alpha}'_{jk\neq j}$ is a "nonlocal tensor" spanning two such coordinate systems. This formalism follows naturally from the linearization of the equations of motion for non-collinear macrospins, and is particularly useful for describing the influence of "tensor damping" on the thermal fluctuations and/or spin-torque critical currents of such multilayer film structures (e.g., as described further in Sec. IV.). Due to the spin-pumping contribution $\vec{\alpha}'^{\text{pump}}_{jk}$, the four individual $\alpha'^{u'v'}_{jk}$ (with $u', v' = x'$ or $y'$) are in general nonzero with $\alpha'^{x'x'}_{jk} \neq \alpha'^{y'y'}_{jk}$ , reflecting the true tensor nature of the damping in this circumstance, which is additionally nonlocal between magnetic layers (i.e., $\alpha'^{u'v'}_{jk\neq j} \neq 0$). The $\alpha'^{u'v'}_{jk}$ are somewhat arbitrary to the extent that one may replace $\tilde{\Re} \leftrightarrow \tilde{\Re} \cdot \Re_2$ in (9), where $\Re_2$ is the $2 \times 2$ matrix representation of any rotation about the $\hat{\boldsymbol{z}}'$ axis.

It is perhaps tempting to contemplate an "inverse linearization" of (9) to obtain a 3D nonlinear Gilbert equation with a fully 3D damping tensor $\vec{\alpha}_{jk} = \Re_j \cdot \vec{\alpha}'_{jk} \cdot \Re_k^{\mathsf{T}}$. However, (9) has a null $\hat{\boldsymbol{z}}'$ component, and contains no information regarding the heretofore undefined quantities $\alpha'^{u'z'}_{jk}$ or $\alpha'^{z'z'}_{jk}$. For local, isotropic/scalar Gilbert damping, one can independently argue on spatial symmetry grounds that $\alpha'^{z'z'}_{\mathrm{G}} = \alpha'^{u'u'}_{\mathrm{G}} = \alpha_{\mathrm{G}}$. However, the analogous extension is not so obviously available for $\vec{\alpha}'^{\text{pump}}_{jk}$, given the intrinsically nonlocal, anisotropic nature of spin-pumping. The proper general equation remains that of (8), with the rightmost term given by that in (5), or its equivalent.

## III. EXAMPLE: FIVE LAYER SYSTEM

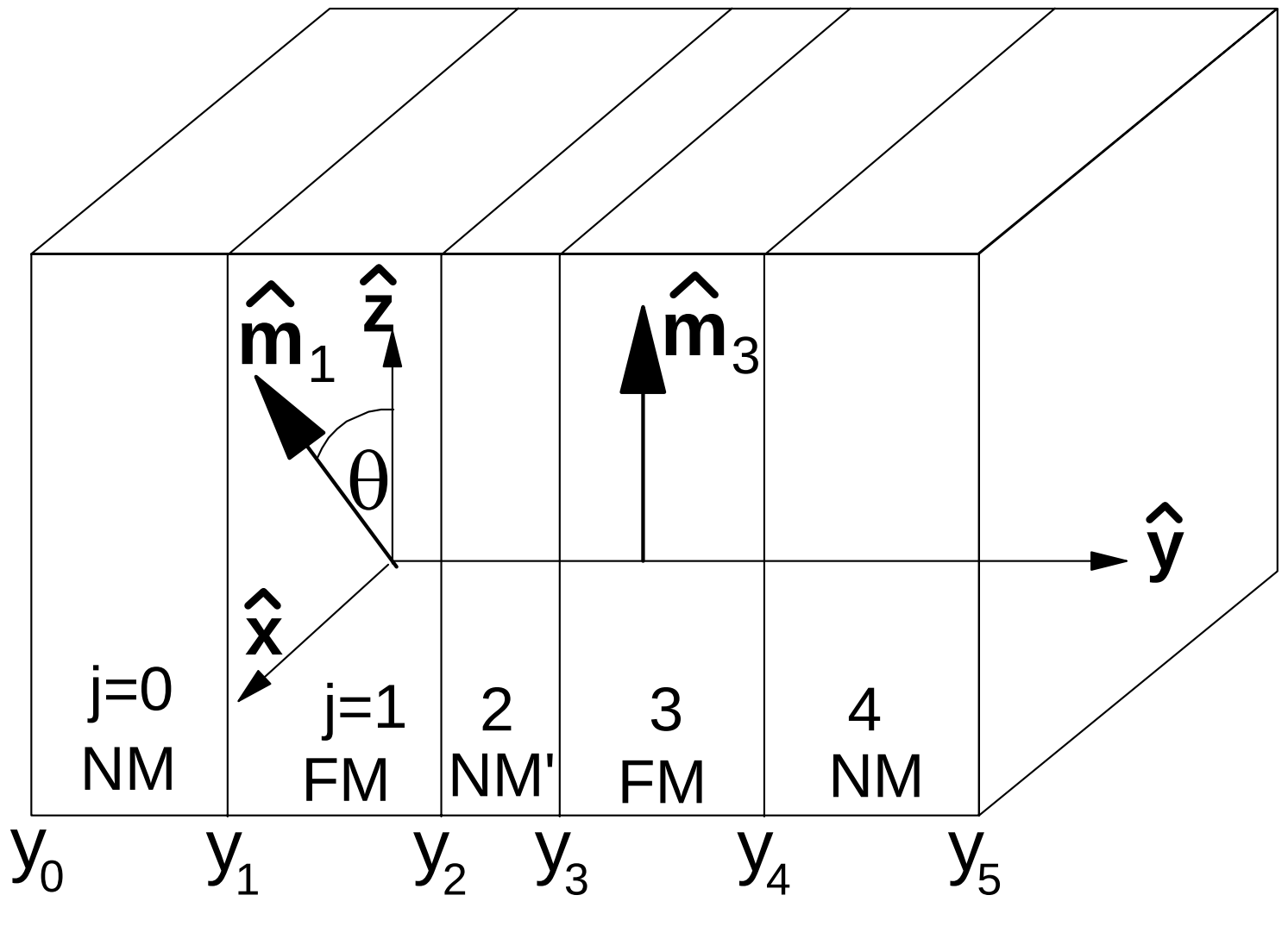

Fig. 2. Cartoon of a prototypical 5-layer CPP-GMR stack (leads not shown) with two FM layers (#1 and #3), sandwiching a central NM spacer layer (#2 ), and with outer NM cap layers (#0 and #4). For discussion purposes described in text, the magnetization vectors $\hat{\boldsymbol{m}}_1$ and $\hat{\boldsymbol{m}}_3$ can be considered to lie in the film plane ($x$-$z$ plane).

Fig. 2 shows a 5-layer system with 2 FM layers resembling a CPP-GMR spin-valve, to be used as a prototype. Although the full generalization is straightforward, the material properties and layer thickness will be assumed symmetric about the central (#2) normal metal spacer layer, which will additionally be taken to have a large spin-diffusion length $l_2 >>> t_2$ (with $t_j$ the thickness of the $j^{\text{th}}$ layer), such that the "ballistic" approximation (B3) applies. The inverse mixing conductances $r_{i=1\text{-}4}^{\uparrow\downarrow}$ will also be assumed to be real. Using the outer boundary conditions described by (B5), one finds for the FM-NM interfaces at $y = y_1$ and $y_4$, that

$$\boldsymbol{J}_{i=1,4}^{\text{NM}} = J_i^{\text{FM}} \hat{\boldsymbol{m}}_{j=1,3} + \frac{r_1^{\uparrow\downarrow} \boldsymbol{J}_i^{\text{pump}}}{[r_1'^{\uparrow\downarrow} \equiv r_1^{\uparrow\downarrow} + (\rho l \text{ hyp}(t/l))_{\text{NM}}]} \tag{10a}$$

$$-\tfrac{1}{2}\Delta V_{i=4}^{\text{FM}} = [r_1' \equiv r_1 + (\rho l \text{ hyp}(t/l))_{\text{NM}}] J_4^{\text{FM}} \tag{10b}$$

where $r_1 = r_4$, $r_1^{\uparrow\downarrow} = r_4^{\uparrow\downarrow}$, and subscript "NM" refers to either outer layer 0 or 4. In (10) and below, $\hat{\boldsymbol{m}}_j \leftrightarrow \hat{\boldsymbol{m}}_{0j}$ are used interchangeably. Inside FM layer 3, (B1,2) have solution expressible as

$$\begin{aligned} &\Delta V_3(y_3 \le y \le y_4) = 2A_3 \sinh((y - y_3)/l_{\text{FM}}) + 2B_3 \cosh((y - y_3)/l_{\text{FM}}) \\ &J_3^{\text{spin}}(y) = 1/(\rho l)_{\text{FM}} [A_3 \cosh((y - y_3)/l_{\text{FM}}) + B_3 \sinh((y - y_3)/l_{\text{FM}}) \\ &B_3 = -[(r_1' + (\rho l \tanh(t/l)_{\text{FM}}]/[(\rho l)_{\text{FM}} + r_1' \tanh(t/l)_{\text{FM}}] A_3 \end{aligned} \tag{11}$$

where the expression for $B_3$ follows from (10b). Subscript "FM" refers to either layers 1 or 3. The boundary conditions (A5) and (A9) applied to the FM/NM boundary at $y = y_3$ yield

$$\tfrac{1}{2}(2B_3 - \Delta \boldsymbol{V}_2) = (r_2 - r_2^{\uparrow\downarrow})[A_3/(\rho l)_{\text{FM}} = \boldsymbol{J}_2^{\text{spin}} \cdot \hat{\boldsymbol{m}}_3]\hat{\boldsymbol{m}}_3 + r_2^{\uparrow\downarrow}(\boldsymbol{J}_2^{\text{spin}} - \boldsymbol{J}_3^{\text{pump}}) \tag{12}$$

where $r_2 = r_3$, $r_2^{\uparrow\downarrow} = r_3^{\uparrow\downarrow}$. The "ballistic" values $\Delta \boldsymbol{V}_2$ and $\boldsymbol{J}_2^{\text{spin}}$ are constant inside central layer 2. Using (11) to eliminate coefficient $B_3$ in (12), the latter may be rewritten as

$$\begin{aligned} &-\tfrac{1}{2}\Delta \boldsymbol{V}_2 = r_2^{\uparrow\downarrow}[(\overleftrightarrow{1} + 2q\, \hat{\boldsymbol{m}}_3 \cdot \hat{\boldsymbol{m}}_3^{\mathrm{T}}) \cdot \boldsymbol{J}_2^{\text{spin}} - \boldsymbol{J}_3^{\text{pump}}] \\ &q \equiv \frac{1}{2r_2^{\uparrow\downarrow}} \left[ r_2 - r_2^{\uparrow\downarrow} + \frac{r_1' + (\rho l \tanh(t/l))_{\text{FM}}}{1 + r_1'[\tanh(t/l)/(\rho l)]_{\text{FM}}} \right] \end{aligned} \tag{13}$$

where $\overleftrightarrow{1}$ is the 3-D identity tensor, and $\hat{\boldsymbol{m}}_3 \cdot \hat{\boldsymbol{m}}_3^{\mathrm{T}}$ denotes the 3-D tensor formed from the vector *outer*-product of $\hat{\boldsymbol{m}}_3$ with itself.

Working through the equivalent computations applied now to the NM/FM interface at $y = y_2$, one finds the analogous result:

$$+\tfrac{1}{2}\Delta\boldsymbol{V}_2 = r_2^{\uparrow\downarrow}[(\overleftrightarrow{1} + 2q\,\hat{\boldsymbol{m}}_1 \cdot \hat{\boldsymbol{m}}_1^{\mathsf{T}}) \cdot \boldsymbol{J}_2^{\text{spin}} - \boldsymbol{J}_2^{\text{pump}}] \tag{14}$$

Eliminating $\Delta\boldsymbol{V}_2$ between (13) and (14) derives the remaining needed result for $\boldsymbol{J}_2^{\text{spin}}$:

$$\cdot\boldsymbol{J}_2^{\text{spin}} = \tfrac{1}{2}\overleftrightarrow{Q} \cdot (\boldsymbol{J}_2^{\text{pump}} + \boldsymbol{J}_3^{\text{pump}}),\ \overleftrightarrow{Q} \equiv [\overleftrightarrow{1} + q\,(\hat{\boldsymbol{m}}_1 \cdot \hat{\boldsymbol{m}}_1^{\mathsf{T}} + \hat{\boldsymbol{m}}_3 \cdot \hat{\boldsymbol{m}}_3^{\mathsf{T}})]^{-1} \tag{15}$$

treating tensor $\overleftrightarrow{Q}$ as the $3 \times 3$ matrix inverse of the [ ]-bracketed tensor in (15). Using (10a) and (15) to compute $J_{i=1\text{-}4}^{\text{NM}}$, then additional use of (4) and (6), allow computation of the $\overleftrightarrow{C}_{jk}$ defined in (7):

$$\begin{gathered}\overleftrightarrow{C}_{11} = \overleftrightarrow{C}_{33} = a\,\overleftrightarrow{1} + b\overleftrightarrow{Q}\,,\ \overleftrightarrow{C}_{13} = \overleftrightarrow{C}_{31} = -b\overleftrightarrow{Q} \\ a \equiv \bar{r}^{\uparrow\downarrow}/r_1'^{\uparrow\downarrow},\ b \equiv \bar{r}^{\uparrow\downarrow}/2r_2^{\uparrow\downarrow};\ 1/\bar{r}^{\uparrow\downarrow} = \tfrac{1}{2}(1/r_1^{\uparrow\downarrow} + 1/r_2^{\uparrow\downarrow})\end{gathered} \tag{16}$$

For explicit evaluation of $\overleftrightarrow{\alpha}'^{\text{pump}}_{jk}$, it is convenient to assume a choice of $\tilde{\Re}_{j=1,3}$ for which $\hat{\boldsymbol{y}}_1' = \hat{\boldsymbol{y}}_3'$, such that $\hat{\boldsymbol{m}}_{03}$ and $\hat{\boldsymbol{m}}_{01}$ lie in the $x'$-$z'$ plane. To simplify the intermediate algebra to obtain $\overleftrightarrow{Q}$ from (15), one can consider "in-plane" magnetizations (Fig.2), taking $\hat{\boldsymbol{m}}_{03} = \hat{\boldsymbol{z}}$, and $\hat{\boldsymbol{m}}_{01}$ in the $x$-$z$ plane ($\hat{\boldsymbol{m}}_{01} \cdot \hat{\boldsymbol{z}} = \cos\theta$). This allows a particularly easy determination of $\tilde{\Re}_j$ for which $\hat{\boldsymbol{y}}_1' = \hat{\boldsymbol{y}}_3' = \hat{\boldsymbol{y}}$:

$$\tilde{\Re}^{\mathsf{T}}_{j=1,3} = \begin{pmatrix} \cos\theta_j & 0 & -\sin\theta_j \\ 0 & 1 & 0 \end{pmatrix};\ \theta_1 = \theta,\ \theta_3 = 0 \tag{17}$$

Using (16) and (17) with (9) allows explicit solution for the $\overleftrightarrow{\alpha}'^{\text{pump}}_{jk}$:

$$\begin{gathered}\overleftrightarrow{\alpha}'^{\text{pump}}_{jk} = \frac{\hbar\gamma}{(4\pi M_s t)_j}\frac{h/2e^2}{\bar{r}_j^{\uparrow\downarrow}}\begin{pmatrix} a\delta_{jk} + b(2\delta_{jk} - 1) & 0 \\ 0 & a\delta_{jk} + bd_{jk} \end{pmatrix} \\ d_{11} = d_{33} = \frac{1 + q + q\cos^2\theta}{1 + 2q + q^2\sin^2\theta},\ d_{13} = d_{31} = \frac{-(1+2q)\cos\theta}{1 + 2q + q^2\sin^2\theta}\end{gathered} \tag{18}$$

Taking $\cos\theta = \hat{\boldsymbol{m}}_{01} \cdot \hat{\boldsymbol{m}}_{03}$, (18) holds for arbitrary orientation of $\hat{\boldsymbol{m}}_{01}$ and $\hat{\boldsymbol{m}}_{03}$, provided the flexibility in choosing the $\tilde{\Re}_{j=1,3}$ is used to maintain $\hat{\boldsymbol{y}}_1' = \hat{\boldsymbol{y}}_3'$. However, for multilayer film stacks with three or more magnetic layers with magnetizations $\hat{\boldsymbol{m}}_{0j}$ that do not all lie in a single plane, it will generally be the case that some of the off-diagonal elements of the $\overleftrightarrow{\alpha}'^{\text{pump}}_{jk}$ will be nonzero.

## IV. DISCUSSION

It is perhaps instructive to compare and contrast the results of (9) and (18). with the prior results of Ref. 3. The latter are for a a trilay*er* stack, corresponding most directly to taking $\rho_{\mathrm{NM}} \to \infty$ in the present model, whereby $\boldsymbol{J}_{i=1,4}^{\mathrm{pump}} = \boldsymbol{J}_{i=1,4}^{\mathrm{NM}} = 0$. It is also effectively equivalent to the 5-layer case with insulating outer boundaries in the limit $(t/l)_{\mathrm{NM}} \to 0$, whereby $\boldsymbol{J}_{i=1,4}^{\mathrm{pump}} \neq 0$ but $\boldsymbol{J}_{i=1,4}^{\mathrm{NM}} \to 0$ due to perfect cancellation by the spin current reflected from the $y_{i=0,5}$ boundaries without intervening spin-flip scattering. Either way, it corresponds to $r_1', r_1'^{\uparrow\downarrow} \to \infty$ in (10) and $a \to 0$ in (16) and (18).

However, a more interesting difference is that Ref. 3 treats $\hat{\boldsymbol{m}}_3$ as stationary (hence $\boldsymbol{J}_3^{\mathrm{pump}} = 0)$, and $\hat{\boldsymbol{m}}_1$ as undergoing a perfectly *circular* precession about $\hat{\boldsymbol{m}}_3$ with a possibly large cone angle $\theta$. By contrast, the present analysis treats $\hat{\boldsymbol{m}}_1$ and $\hat{\boldsymbol{m}}_3$ equally as quasi-stationary vectors which undergo small but otherwise random fluctuations about their equilibrium positions $\hat{\boldsymbol{m}}_{01}$ and $\hat{\boldsymbol{m}}_{03}$, with $\hat{\boldsymbol{m}}_{03} \cdot \hat{\boldsymbol{m}}_{03} = \cos\theta$. To further elucidate this distinction, one can assume the aforementioned physical model of Ref. 3, and reanalyze that situation in terms of the present formalism. With $d\boldsymbol{m}_3/dt = 0 = \boldsymbol{J}_3^{\mathrm{pump}}$, and by explicitly inserting the condition (e.g., from (3)) that $\boldsymbol{J}_2^{\mathrm{pump}} \cdot \hat{\boldsymbol{m}}_1 \equiv 0$, an explicit solution of (15) can be expressed in the form:

$$\boldsymbol{J}_2^{\mathrm{NM}} = \tfrac{1}{2}[\boldsymbol{J}_2^{\mathrm{pump}} + \frac{q^2\cos\theta\,\hat{\boldsymbol{m}}_1 - q(q+1)\hat{\boldsymbol{m}}_3}{(1+q)^2 - q^2\cos^2\theta}\boldsymbol{J}_2^{\mathrm{pump}} \cdot \hat{\boldsymbol{m}}_3] \tag{19}$$

Combining (19) with the earlier result from (5) and then (3) (with $\varepsilon = 0)$, it is readily found that

$$\begin{aligned}
\hat{\boldsymbol{m}}_1 \times \frac{d\hat{\boldsymbol{m}}_1}{dt} &\Leftrightarrow \frac{\gamma}{(M_s t)_1}\hat{\boldsymbol{m}}_1 \times \frac{1}{A}\frac{d\boldsymbol{S}_1}{dt} = -\frac{\hbar}{2e}\frac{\gamma}{(M_s t)_1}\hat{\boldsymbol{m}}_1 \times \boldsymbol{J}_2^{\mathrm{NM}} \\
&= -\frac{\hbar}{4e}\frac{\gamma}{(M_s t)_1}\left(\hat{\boldsymbol{m}}_1 \times \boldsymbol{J}_2^{\mathrm{pump}} + \frac{q(q+1)(\hat{\boldsymbol{m}}_3 \cdot \boldsymbol{J}_2^{\mathrm{pump}})}{(1+q)^2 - q^2\cos^2\theta}\hat{\boldsymbol{m}}_3 \times \hat{\boldsymbol{m}}_1\right) \\
&= -\left[\frac{\hbar\gamma}{(8\pi M_s t)_1}\frac{h/2e^2}{r_2^{\uparrow\downarrow}}\left(1 - \frac{q(q+1)\sin^2\theta}{(1+q)^2 - q^2\cos^2\theta}\right)\right]\frac{d\hat{\boldsymbol{m}}_1}{dt}
\end{aligned} \tag{20}$$

The last result in (20) uses $\boldsymbol{J}_2^{\text{pump}}$ from (3), and the fact that $|\hat{\boldsymbol{m}}_3 \times \hat{\boldsymbol{m}}_1| = \sin\theta$, and that $d\hat{\boldsymbol{m}}_1 / dt$ and $\hat{\boldsymbol{m}}_3 \times \hat{\boldsymbol{m}}_1$ are *parallel* vectors in the case of steady *circular* precession of $\hat{\boldsymbol{m}}_1$ about a fixed $\hat{\boldsymbol{m}}_3$. It is the direct equivalent of Eq. (9) of Ref. 3 with the identification $\nu \Leftrightarrow q/(q+1)$.

Although the final expression in (20) is azimuthally invariant with vector orientation of $\hat{\boldsymbol{m}}_1$, it is most convenient to compare it with (18) at that instant where $\hat{\boldsymbol{m}}_1$ is "in-plane" as shown in Fig. 2. At that orientation, $d\hat{\boldsymbol{m}}_1 / dt \to dm_{1y} / dt = dm'_{1y} / dt$, and it is immediately confirmed from (9) and (18) (with $a \to 0$) that the [ ]-term in (20) is simply the tensor element $\alpha_{11}'^{y'y'}$ of $\ddot{\alpha}_{11}'^{\text{pump}}$. It is now seen that the analysis of Ref. 3 happens to mask the tensor nature of the spin-pump damping by its restricting attention a specific form of the motion of the magnetization vectors, which in this case singles out the single diagonal element of the $\ddot{\alpha}_{11}'^{\text{pump}}$ tensor along the axis perpendicular to the plane formed by vectors $\hat{\boldsymbol{m}}_1$ and $\hat{\boldsymbol{m}}_3$. The very recent results of Ref. 6 do address this deficiency of generality, and reveal the tensor nature of $\ddot{\alpha}_{11}'^{\text{pump}}$ with specific results for $\theta = 0, \pi/2,$ and $\pi$. The present Sec. III additionally includes the nonlocal tensors $\ddot{\alpha}_{13}'^{\text{pump}} = \ddot{\alpha}_{31}'^{\text{pump}}$, as well as diagonal terms $a\delta_{jk}$ in (18) (and the variation in parameter $q$) when it is *not* the case that $r_{\text{NM-FM}} \ll (\rho l)_{\text{NM}} \,\text{hyp}(t_{\text{NM}} / l_{\text{NM}})$ in boundary condition (B4). The latter condition will likely apply in the case of the technological important example of CPP-GMR spin-valves.

Speaking of such, two important practical issues for these devices involve thermal magnetic noise and spin-torque induced oscillations. As described previously[8], an explicit linearization of the $\boldsymbol{H}^{\text{eff}}$ term in (9) about equilibrium state $\hat{\boldsymbol{m}}_0$ that is a minimum of the free energy $E$ leads to the following matrix form of the linearized Gilbert equation including spin-pumping (with $J_e = 0$):

$$
\begin{aligned}
&\sum_k (\ddot{D}'_{jk} + \ddot{G}'_{jk}) \cdot \frac{d\boldsymbol{m}'_k}{dt} + \sum_k \ddot{H}'_{jk} \cdot \boldsymbol{m}'_k \cdot = \boldsymbol{h}'_j(t) \equiv p_j \tilde{\Re}_j^{\mathsf{T}} \cdot \boldsymbol{h}_j(t) \\
&\ddot{D}'_{jk} \equiv \frac{p_j \ddot{\alpha}'_{jk} + p_k \ddot{\alpha}'_{kj}}{2\gamma}, \; \ddot{G}'_{jk} \equiv \left[ \begin{pmatrix} 0 & -1 \\ 1 & 0 \end{pmatrix} \frac{p_j}{\gamma} \delta_{jk} + \frac{p_j \ddot{\alpha}'_{jk} - p_k \ddot{\alpha}'_{kj}}{2\gamma} \right] \\
&\ddot{H}_{jk} \equiv (\hat{\boldsymbol{m}}_{0j} \cdot \boldsymbol{H}_{0j}^{\text{eff}}) \ddot{1} \delta_{jk} - \frac{\partial \boldsymbol{H}_{0j}^{\text{eff}}}{\partial \hat{\boldsymbol{m}}_k}, \; \ddot{H}'_{jk} \equiv \tilde{\Re}_j^{\mathsf{T}} \cdot \ddot{H}_{jk} \cdot \tilde{\Re}_k \\
&\boldsymbol{H}_{0j}^{\text{eff}} \equiv \frac{-1}{\Delta m} \frac{\partial E(\hat{\boldsymbol{m}}_0)}{\partial \hat{\boldsymbol{m}}_j}, \; p_j \equiv \frac{(M_s t A)_j}{\Delta m}
\end{aligned}
\qquad (21)
$$

where the $\boldsymbol{h}_j(t)$ are small perturbation fields. The form of $\vec{D}'_{jk}$ and $\vec{G}'_{jk}$ in (21) is chosen so that they retain the original delineation[8] as symmetric and antisymmetic tensors regardless of the symmetry of $\vec{\alpha}'_{jk}$. By use of a *fixed* "reference-moment" $\Delta m$ in the definition of $\boldsymbol{H}_j^{\text{eff}}$, the "stiffness-field" tensor-matrix $H'^{u'v'}_{jk} \propto \partial E / \partial m'_{ju'} \partial m'_{kv'}$ is symmetric positive-definite, and $\delta E = -A \sum_j (M_s t)_j \boldsymbol{h}_j \cdot \delta \boldsymbol{m}_j$ $= -\Delta m \sum_j \boldsymbol{h}'_j \cdot \boldsymbol{m}'_j$ has the proper conjugate form so that (21) are now ready to directly apply fluctuation-dissipation expressions specifically suited to such linear matrix equations of motion.[8] Treating the fields $\boldsymbol{h}'_j(t)$ now as thermal fluctuation fields driving the $\boldsymbol{m}'_j(t)$-fluctuations,

$$\langle h'_{ju'}(\tau) h'_{kv'}(0) \rangle = \frac{2k_B T}{\Delta m} D'^{u'v'}_{jk} \delta(\tau) \Leftrightarrow S'_{h'_{ju'}h'_{kv'}}(\omega) = \frac{2k_B T}{\Delta m} D'^{u'v'}_{jk} \tag{22}$$

are the time-correlation or cross power spectral density (PSD) Fourier transform pairs. Through their relationship described in (21), the nonlocal, tensor nature of the spin-pumping contribution $\vec{\alpha}'^{\text{pump}}_{jk}$ to $\vec{\alpha}'_{jk}$ is directly translated into those of the $2N_{\text{FM}} \times 2N_{\text{FM}}$ system "damping tensor-matrix" $\vec{\boldsymbol{D}}' \leftrightarrow D'^{u'v'}_{jk}$, where $N_{\text{FM}}$ is the number of FM layers in the multilayer film stack. The cross-PSD tensor-matrix $\vec{\boldsymbol{S}}'_{\boldsymbol{m}'\boldsymbol{m}'}(\omega) \leftrightarrow S'_{m'_{ju'}m_{kv'}}(\omega)$ for the $\boldsymbol{m}'$-fluctuations can then be expressed as[8]

$$\begin{gathered}\vec{\boldsymbol{S}}'_{\boldsymbol{m}'\boldsymbol{m}'}(\omega) = \frac{k_B T}{i\omega \Delta m}[\vec{\chi}'(\omega) - \vec{\chi}'^{\dagger}(\omega)] \;\rightarrow\; \vec{\chi}'(\omega) \cdot \vec{\boldsymbol{S}}'_{\boldsymbol{h}'\boldsymbol{h}'} \cdot \vec{\chi}'^{\dagger}(\omega) \\ \vec{\chi}'(\omega) \equiv [\vec{\boldsymbol{H}}' - i\omega(\vec{\boldsymbol{G}}' + \vec{\boldsymbol{D}}')]^{-1}\end{gathered} \tag{23}$$

where $\vec{\chi}'(\omega)$ is the complex susceptibility tensor-matrix for the $\{\boldsymbol{m}', \boldsymbol{h}'\}$ system, and $\vec{\chi}'^{\dagger}(\omega)$ its Hermitian transpose. It has been theoretically argued[10] that (22), and thus the *second* expression in (23), remain valid when $J_e \neq 0$, despite spin-torque contributions to $\boldsymbol{H}_j^{\text{eff}}$ resulting in an *asymmetric* $\vec{\boldsymbol{H}}'$ (e.g., see (25)) that violates the condition of thermal equilibrium implicitly assumed for the fluctuation dissipation relations.

Since $\vec{\boldsymbol{H}}'$ is in general a fully nonlocal with anisotropic/tensor character, any additional tensor nature of $\vec{\boldsymbol{D}}$ will likely be altered or muted as to the influence on the detectable $\boldsymbol{m}'$-fluctuations. As an example, one can again consider the situation depicted in Fig. 2, applied to the case of a CPP-GMR spin-valve with typical *in-plane* magnetization. The device's output noise PSD will reflect fluctuations in $\hat{\boldsymbol{m}}_1 \cdot \hat{\boldsymbol{m}}_3$. Taking $\hat{\boldsymbol{m}}_3$ to again play the simplifying role of an ideal fixed (or pinned) reference layer (i.e., $d\hat{\boldsymbol{m}}_3 / dt \to 0$ ), the PSD will be proportional to $\sin^2 \theta S'_{m'_{1x'} m'_{1x}}(\omega)$ . As was also shown previously,[11] it follows from (23) (and assuming azimuthal symmetry $H'^{x'y'}_{11} = H'^{y'x'}_{11} = 0$) that

$$S'_{m'_{1x'} m'_{1x}}(\omega) \cong \frac{2k_B T\gamma}{(M_s tA)_1} \frac{\alpha'^{x'x'}_{11}(H'^{y'y'}_{11} / H'^{x'x'}_{11})\,\omega_0^2 + \alpha'^{y'y'}_{11}\omega^2}{(\omega^2 - \omega_0^2)^2 + (\omega\Delta\omega)^2}$$
$$\omega_0 = \gamma\sqrt{H'^{y'y'}_{11} H'^{x'x'}_{11}} \quad \Delta\omega = \alpha'^{x'x'}_{11} H'^{y'y'}_{11} + \alpha'^{y'y'}_{11} H'^{x'x'}_{11} \tag{24}$$

treating $\alpha'^{x'x'}_{11}\alpha'^{y'y'}_{11} \lll 1$. The tensor influence of the $\alpha'^{u'u'}_{11}$ is seen to be weighted by the relative size of the stiffness-field matrix elements $H'^{v'v'}_{11}$. For the thin film geometries $t \ll \sqrt{A}$ typical of such devices, out-of-plane demagnetization field contribution typically result in $H'^{y'y'}_{11}$ that are an order of magnitude larger than $H'^{x'x'}_{11}$. Since $\alpha'^{y'y'}_{11} \le \alpha'^{x'x'}_{11}$ from (18), it follows that the linewidth $\Delta\omega$ and the PSD $S'_{m'_{1x'} m'_{1x}}(\omega \le \omega_0)$ in the spectral range of practical interest will both be expected to be determined primarily by $\alpha'^{x'x'}_{11}$.

A similar circumstance also applies to the important problem of critical currents for spin-torque magnetization excitation in CPP-GMR spin valves with $J_e \ne 0$. Consider the same example as above, again treating $\hat{\boldsymbol{m}}_3$ as stationary, and seeking nontrivial solutions of (21) (with $\boldsymbol{h}'(t) = 0$) of the form $\boldsymbol{m}'_1(t) \propto e^{-st}$. Summarizing results obtainable from (5), (8), and (21)

$$\boldsymbol{H}^{\mathrm{eff}}_1 = \boldsymbol{H}^{\mathrm{eff}}_1\Big|_{J_e=0} + \frac{\hbar / 2e}{(M_s t)_1}\boldsymbol{J}^{\mathrm{NM}}_2 \times \hat{\boldsymbol{m}}_1$$
$$\det\begin{pmatrix} H'^{x'x'}_{11} - s'\alpha'^{x'x'}_{11} & H'^{x'y'}_{11} + s' \\ H'^{y'x'}_{11} - s' & H'^{y'y'}_{11} - s'\alpha'^{y'y'}_{11} \end{pmatrix} = 0 \tag{25}$$

where $s' = s/\gamma$, $\alpha'^{u'v'}_{11}$ as in (18),and where $\boldsymbol{J}^{\mathrm{NM}}_2 \propto J_e$ in (25) is now the solution of the transport equations with $\boldsymbol{J}^{\mathrm{pump}} = 0$ but $J_e \ne 0$. The cross-product form of the spin-torque contribution to $\boldsymbol{H}^{\mathrm{eff}}_1$ explicitly yields an asymmetric/nonreciprocal contribution $H'^{x'y'}_{11} - H'^{y'x'}_{11} \propto J_e$ to $\overleftrightarrow{\boldsymbol{H}}'$. The critical current density is that value of $J_e$ where $\mathrm{Re}\, s$ becomes negative. Given the basic stability criterion that $\det \overleftrightarrow{H}'_{11} > 0$, the spin-torque critical condition from (25) can be expressed as

$$\alpha'^{y'y'}_{11} H'^{x'x'}_{11} + \alpha'^{x'x'}_{11} H'^{y'y'}_{11} = H'^{x'y'}_{11} - H'^{y'x'}_{11} \tag{26}$$

Like for thermal noise, the spin-torque critical point should again be determined primarily by $\alpha'^{x'x'}_{11}$ for in-plane magnetized CPP-GMR spin-valves with typical $H'^{y'y'}_{11} >> H'^{x'x'}_{11}$. This simply reflects the fact that the (quasi-uniform) modes of thermal fluctuation or critical-point spin-torque oscillation tend to exhibit rather "elliptical", mostly in-plane motion when $H'^{y'y'}_{11} >> H'^{x'x'}_{11}$. This is obviously different than the steady, pure circular precession described in Ref. 3, which contrastingly highlights the influence of $\alpha'^{y'y'}_{11}$, along with its interesting, additional $\theta$-dependence.

## APPENDIX A

The well known "circuit theory" formulation[12] of the boundary conditions for the electron charge current density $J_e$ and the (dimensionally equivalent) spin current density $\boldsymbol{J}^{\text{spin}}_{\text{NM}}$ at a FM/NM interface can (taking $\Delta\boldsymbol{V}_{\text{FM}} = \Delta V_{\text{FM}}\hat{\boldsymbol{m}}$) be expressed as

$$J_e = (G^{\uparrow} + G^{\downarrow})(\overline{V}_{\text{NM}} - \overline{V}_{\text{FM}}) + \tfrac{1}{2}(G^{\uparrow} - G^{\downarrow})(\Delta\boldsymbol{V}_{\text{NM}} \cdot \hat{\boldsymbol{m}} - \Delta V_{\text{FM}}) \tag{A1}$$

$$\begin{aligned} \boldsymbol{J}^{\text{spin}}_{\text{NM}} &= [(G^{\uparrow} - G^{\downarrow})(\overline{V}_{\text{NM}} - \overline{V}_{\text{FM}}) + \tfrac{1}{2}(G^{\uparrow} + G^{\downarrow})(\Delta\boldsymbol{V}_{\text{NM}} \cdot \hat{\boldsymbol{m}} - \Delta V_{\text{FM}})]\hat{\boldsymbol{m}} \\ &+ \operatorname{Re} G^{\uparrow\downarrow}(\hat{\boldsymbol{m}} \times \Delta\boldsymbol{V}_{\text{NM}} \times \hat{\boldsymbol{m}}) + \operatorname{Im} G^{\uparrow\downarrow}(\Delta\boldsymbol{V}_{\text{NM}} \times \hat{\boldsymbol{m}}) \end{aligned} \tag{A2}$$

in terms of spin-independent electric potential $\overline{V}$ and accumulation $\Delta\boldsymbol{V}$ $(= e\Delta\mu)$. Setting $J_e = 0$ in (A1) and substituting into (A2), one obtains in the limit $\operatorname{Im} G^{\uparrow\downarrow} \to 0$ the result

$$\boldsymbol{J}^{\text{spin}}_{\text{NM}}\Big|_{J_e=0} = \frac{2G^{\uparrow}G^{\downarrow}}{G^{\uparrow} + G^{\downarrow}}(\Delta\boldsymbol{V}_{\text{NM}} \cdot \hat{\boldsymbol{m}} - \Delta V_{\text{FM}})\hat{\boldsymbol{m}} + G^{\uparrow\downarrow}(\hat{\boldsymbol{m}} \times \Delta\boldsymbol{V}_{\text{NM}} \times \hat{\boldsymbol{m}}) \tag{A3}$$

Comparing with Eq. (4) of Tserkovnyak et al.[3] (with $\Delta\boldsymbol{V} \Leftrightarrow \mu_s$) and remembering the present conversion of $\boldsymbol{J}^{\text{spin}}_{\text{NM}} \leftrightarrow -(A\hbar/2e)^{-1}\boldsymbol{I}^{\text{spin}}_{\text{NM}}$, one immediately makes the identification

$$g^{\uparrow\downarrow} = 2A(h/2e^2)G^{\uparrow\downarrow} \tag{A4}$$

relating dimensionless $g^{\uparrow\downarrow}$ in (1) to $G^{\uparrow\downarrow}$, the conventional mixing conductance (per area).

The common approximations that $\boldsymbol{J}^{\text{spin}}_{\text{FM}} = J^{\text{spin}}_{\text{FM}}\hat{\boldsymbol{m}}$ inside all FM layers, and that *longitudinal* spin current density is conserved at FM/NM interfaces, yields the usual interface boundary condition

$$\boldsymbol{J}_{\text{NM}}^{\text{spin}} \cdot \hat{\boldsymbol{m}} = J_{\text{FM}}^{\text{spin}} \tag{A5}$$

Solving for $\boldsymbol{J}_{\text{NM}}^{\text{spin}} \cdot \hat{\boldsymbol{m}}$ from (A2) then leads (with (A1)) to a second scalar boundary condition:

$$\overline{V}_{\text{NM}} - \overline{V}_{\text{FM}} = \frac{G^{\uparrow} + G^{\downarrow}}{4G^{\uparrow}G^{\downarrow}} J_e - \frac{G^{\uparrow} - G^{\downarrow}}{4G^{\uparrow}G^{\downarrow}} J_{\text{FM}}^{\text{spin}} \tag{A6}$$

Equation (A6) is identical in form with the standard (collinear) Valet-Fert model,[7] and immediately yields the following identifications

$$r = \frac{G^{\uparrow} + G^{\downarrow}}{4G^{\uparrow}G^{\downarrow}}, \quad \gamma = \frac{G^{\uparrow} - G^{\downarrow}}{G^{\uparrow} + G^{\downarrow}} \tag{A7}$$

for the conventional Valet-Fert interface parameters $r$ and $\gamma$.

The three vector terms on the right of (A2) are mutually orthogonal. Working in a rotated (primed) coordinate system where $\hat{\boldsymbol{z}}' = \hat{\boldsymbol{m}}'$, (A1) and (A2) can be similarly inverted to solve for the three components of the vector $(\Delta\boldsymbol{V}'_{\text{NM}} - \Delta V_{\text{FM}}\hat{\boldsymbol{m}}')$ in terms of $\boldsymbol{J}'^{\text{spin}}_{\text{NM}}$, $J_{\text{FM}}^{\text{spin}}$, and $J_e$. A final transformation back to the original (umprimed) coordinates yields the vector interface-boundary condition

$$\begin{gathered}\tfrac{1}{2}(\Delta\boldsymbol{V}_{\text{NM}} - \Delta V_{\text{FM}}\hat{\boldsymbol{m}}) = [(r - \operatorname{Re} r^{\uparrow\downarrow})J_{\text{FM}}^{\text{spin}} - r\gamma J_e]\hat{\boldsymbol{m}} + \operatorname{Re} r^{\uparrow\downarrow}\boldsymbol{J}_{\text{NM}}^{\text{spin}} + \operatorname{Im} r^{\uparrow\downarrow}\hat{\boldsymbol{m}} \times \boldsymbol{J}_{\text{NM}}^{\text{spin}} \\ r^{\uparrow\downarrow} \equiv 1/(2G^{\uparrow\downarrow}) = (h/2e^2)/(g^{\uparrow\downarrow}/A)\end{gathered} \tag{A8}$$

Combined with (A4), the last relation in (A8) yields (2). Equation (A8) is a generalization of Valet-Fert to the non-collinear case.

As noted by Tserkovnyak et al.,[3] boundary conditions (A3) do not directly include spin-pumping terms, but instead involve only "backflow" terms $\boldsymbol{J}_{\text{NM}}^{\text{spin}} \leftrightarrow \boldsymbol{J}_{\text{NM}}^{\text{back}}$ in the NM layer. With spin-pumping physically present, $\boldsymbol{J}_{\text{NM}}^{\text{back}}$ arises as the response to the spin accumulation $\Delta\boldsymbol{V}_{\text{NM}}$ created by $\boldsymbol{J}^{\text{pump}}$. It follows that $\boldsymbol{J}_{\text{NM}}^{\text{back}} = \boldsymbol{J}_{\text{NM}}^{\text{spin}} - \boldsymbol{J}^{\text{pump}}$, where $\boldsymbol{J}_{\text{NM}}^{\text{spin}}$ is henceforth the *total* spin current in the NM layer. Thus, including spin-pumping in Valet-Fert transport equations is then a matter of replacing $\boldsymbol{J}_{\text{NM}}^{\text{spin}} \to \boldsymbol{J}_{\text{NM}}^{\text{spin}} - \boldsymbol{J}^{\text{pump}}$ in (A8). The modified form of (A8), for a FM/NM interface, becomes:

$$\begin{aligned}\tfrac{1}{2}(\Delta\boldsymbol{V}_{\text{NM}} - \Delta V_{\text{FM}}\hat{\boldsymbol{m}}) &= [(r - \operatorname{Re} r^{\uparrow\downarrow})J_{\text{FM}}^{\text{spin}} - r\gamma J_e]\hat{\boldsymbol{m}} \\ &\quad + \operatorname{Re} r^{\uparrow\downarrow}(\boldsymbol{J}_{\text{NM}}^{\text{spin}} - \boldsymbol{J}^{\text{pump}}) + \operatorname{Im} r^{\uparrow\downarrow}\hat{\boldsymbol{m}} \times (\boldsymbol{J}_{\text{NM}}^{\text{spin}} - \boldsymbol{J}^{\text{pump}})\end{aligned} \tag{A9}$$

For an NM/FM interface, the sign is flipped on the left sides of (A6) and (A9).

## APPENDIX B

For 1-D transport (flow along the $y$-axis), the quasi-static Valet-Fert[7] (drift-diffusion, quasi-static) transport equations can be written as[9]

$$\frac{\partial^2 \Delta\boldsymbol{V}}{\partial y^2} = \frac{\Delta\boldsymbol{V}}{l^2}\,,\ \frac{\partial}{\partial y}\left[J_e = \frac{1}{\rho}\left(\frac{\partial \overline{V}}{\partial y} + \frac{1}{2}\beta\hat{\boldsymbol{m}}\cdot\frac{\partial \Delta\boldsymbol{V}}{\partial y}\right)\right] = 0$$
$$\text{along with } \boldsymbol{J}^{\text{spin}} = \frac{1}{\rho}\left(\beta\frac{\partial\overline{V}}{\partial y}\hat{\boldsymbol{m}} + \frac{1}{2}\frac{\partial\Delta\boldsymbol{V}}{\partial y}\right) \tag{B1}$$

where $\rho$ = bulk resistivity[13], $l$ = spin diffusion length, and $\beta$ = bulk/equilibrium spin current polarization in FM layers ($\beta \equiv 0$ in NM layers). The solution for any one layer has the form

$$\overline{V} = \rho J_e y + C - \tfrac{1}{2}\beta\Delta\boldsymbol{V}\cdot\hat{\boldsymbol{m}}\,,\ \Delta\boldsymbol{V} = \boldsymbol{A}e^{y/l} + \boldsymbol{B}e^{-y/l}$$
$$\text{for FM layers: } \boldsymbol{A} = A\hat{\boldsymbol{m}}\,,\ \boldsymbol{B} = B\hat{\boldsymbol{m}} \tag{B2}$$

In the case where $l >>>$ film thickness, one may employ an alternative "ballistic" approximation:

$$\Delta\boldsymbol{V} = \boldsymbol{A}\,,\ \boldsymbol{J}^{\text{spin}} = \boldsymbol{B}\,,\ \overline{V} = C \tag{B3}$$

It is not necessary to solve for the $\overline{V}$ and/or the C-coefficients using (A6) if only $\Delta\boldsymbol{V}$ and $\boldsymbol{J}^{\text{spin}}$ are required. The remaining coefficients are determined by the interface boundary conditions (A5), (A6,7) and (A9), and external boundary conditions at the outer two surfaces of the film stack.

Regarding the latter, one approximation is to treat the external "leads" (with quasi-infinite cross section) as equilibrium reservoirs and set $\Delta\boldsymbol{V}(y = y_{i=0,N}) \to 0$ at the outermost ($i$=0, $N$) lead-stack interfaces of an $N$-layer stack (Fig. 1). The complimentary approximation is of an insulating boundary, with .$\boldsymbol{J}^{\text{spin}}(y = y_{i=0,N}) \to 0$. For the case (such as in Sec. III) where the outer ($j$=0, $N$-1) layers are NM, and the adjacent inner ($j$=1, $N$-2) layers are FM, it is readily found using (B1) and (B2) that

$$\Delta\boldsymbol{V}^{\text{NM}}_{i=1,N-1} = \pm 2(\rho l)_{j=0,N-1}\,\text{hyp}(t_j/l_j)\boldsymbol{J}^{\text{NM}}_i \tag{B4}$$

where hyp( ) = tanh( ) or coth( ) for equipotential, or insulating boundaries, respectively. Combining (B4) with (A9), and neglecting $\text{Im}\,r^{\uparrow\downarrow}$, one finds for $J_e = 0$ that

$$\pm \tfrac{1}{2}\Delta V_{i=1,N-1}^{\mathrm{FM}} = [r_i + (\rho l)_{j=0,N-1}\,\mathrm{hyp}(t_j / l_j)]J_i^{\mathrm{FM}}$$

$$\boldsymbol{J}_i^{\mathrm{NM}} = J_i^{\mathrm{FM}}\hat{\boldsymbol{m}} + \frac{r_i^{\uparrow\downarrow}\boldsymbol{J}_i^{\mathrm{pump}}}{r_i^{\uparrow\downarrow} + (\rho l)_j\,\mathrm{hyp}(t_j / l_j)} \qquad \text{(B5)}$$

ACKNOWLEDGMENT

The author would like to thank Y. Tserkovnyak for bringing Ref. 6 to his attention.